\documentclass[twocolumn,secnumarabic,amssymb, nobibnotes, aps]{revtex4-1}

\usepackage{graphicx}
\usepackage{dcolumn}
\usepackage{bm}
\usepackage{amsmath}

\usepackage{color}
\usepackage{braket}
\usepackage{physics}

\newcommand{\rr}{{\bf r}}

\begin{document}

\begin{abstract}
We introduce a model for an artificial neuron which is based on ballistic transport in a multi-terminal device. Unlike standard configurations, the proposed design embeds the synaptic weights into the active region, thus significantly reducing the complexity of the input terminals. This is achieved by defining the basic elements of the ballistic artificial neuron as follows: the input values are set by the incoming wavefunctions amplitudes, while the weights correspond to the scattering matrix elements. Furthermore, the output value of the activation function of the artificial neuron is given by the transmission function. By tuning the gate voltage, the scattering potential and, consequently, the weights are changed so that the value of the transmission function gets closer to the target output, which is essential in the training process of artificial neural networks. Thus, we provide here the {\it modus operandi} of a ballistic artificial neuron.
\end{abstract}

\title{A ballistic transport model for an artificial neuron}

\author{George Alexandru Nemnes$^{1,2,3*}$ and Daniela Dragoman$^{1,4}$}
\affiliation{$^1$University of Bucharest, Faculty of Physics, MDEO Research Center, 077125 Magurele-Ilfov, Romania}%
\email{E-mail: nemnes@solid.fizica.unibuc.ro}
\affiliation{$^2$Horia Hulubei National Institute for Physics and Nuclear Engineering, 077126 Magurele-Ilfov, Romania}%
\affiliation{$^3$Research Institute of the University of Bucharest (ICUB), Mihail Kogalniceanu Blvd 36-46, 050107 Bucharest, Romania}%
\affiliation{$^4$Academy of Romanian Scientists, Splaiul Independentei 54, Bucharest 050094, Romania}%

\maketitle

\section{Introduction}

Presently, artificial intelligence (AI) applications are well integrated in our everyday lives, ranging from medical imaging and diagnosis, robotics, stock markets, pattern recognition, security and many others. An important subdivision of AI is represented by the machine learning techniques which consists of a variety of statistical methods and algorithms. In particular, artificial neural networks (ANNs) have proved successful in rather different fields, such as classification problems, signal processing or screening of new materials. Although ANNs are typically emulated by software packages with a certain degree of flexibility, for an optimal performance, an increasingly higher number of applications, particularly real time data acquisition, demand specialized modules in form of hardware acceleration systems for the ANN models or even a full hardware implementation. The latter requires an efficient design of artificial neurons (ANs) as main building blocks. Traditionally, FPGA systems have been used to implement ANNs \cite{334585}. 

There are actually many AN models that have been considered and several hardware synaptic implementations have been developed with current technologies \cite{schuman2017survey}. Most AN models are, to some degree, biologically inspired. In biologically-plausible models they closely mimic the biological functions like in the Hodgkin-Huxley model \cite{doi:10.1113/jphysiol.1952.sp004764}, in the simplified Moris-Lecar model \cite{MORRIS1981193} and in other more recent implementations \cite{Yi2018}. In biologically-inspired models, the spiking behavior of the neuron is captured, with the advantage of a less sophisticated hardware implementation, like in the Izhikevich model \cite{1257420}. Most ANNs currently use McCulloch-Pitts neurons \cite{McCulloch1943} with several types of activation functions.

One recurring element in the neuromorphic systems is the synapse implementation and this aspect is critical for an efficient ANN design. Traditional synaptic devices have been implemented with SRAM and DRAM, but these are not optimal in terms of cost, scalability and power consumption. More recently, inspired by biological plasticity mechanisms, the memristors have become an important alternative \cite{Jeong_2018}, due to their non-volatility, fast-switching and for the possibility to achieve high density. The memristor devices have been considered for weight storage, e.g. in McCulloch-Pitts type neurons.

In the past few years the typical device lengths have decreased and the quantum transport became more and more relevant. In this paper we introduce a ballistic artificial neuron (BAN) model, which uses coherent electron transport to process input signals in the fashion of a McCulloch-Pitts type neuron. Models of artificial quantum neurons have been considered before, e.g. using the superposition of wavefunctions and linear combinations of basis elements \cite{10.1007/978-3-7091-6492-1_106}. An important feature of the present BAN model is the inclusion of the synaptic weights into the device active region, which simplifies considerably the design of the synapses. This is achieved by associating the synaptic weights with scattering matrix elements, which depend on the potential inside the device region. In this way the leads corresponding to the in-let ports have the functionality of simple ballistic conductors, which support the coherent propagation of the incoming wave-functions. The weights are controlled by an external gate voltage so that the desired output, i.e. the transmission function, can be properly corrected during the training process. Applying the activation function is here equivalent to measuring the transmission function in the out-let port. It is also worth noting that the proposed compact design ensures not only high-density, but also allows for in-memory computing.

The BAN model is analyzed in the framework of multi-terminal, multi-channel scattering formalism using the R-matrix method \cite{PhysRev.72.29,SMRCKA1990221,PhysRevB.63.085319,PhysRevB.58.16209}. This approach was applied to describe the coherent transport in ballistic nanotransistors \cite{doi:10.1063/1.1748858,doi:10.1063/1.2113413,doi:10.1063/1.3269704}, spin transport in the presence of spin-orbit coupling  \cite{1742-6596-338-1-012012}, low temperature thermoelectric effects in nanowires \cite{NEMNES20101613} and the propagation of wavepackets across interfaces \cite{NEMNES2016109}. Furthermore, snaking states in core-shell nanowires were investigated theoretically \cite{PhysRevB.93.205445} and the results were suitably correlated with experiment \cite{doi:10.1021/acs.nanolett.6b01840}. More recently, the R-matrix method has been employed to describe the functionality of a reconfigurable quantum logic gate \cite{NEMNES201913}.
 
The paper is organized as follows. In the following section, the BAN device model is introduced and the R-matrix formalism is presented, explicitly indicating the connection between the scattering S-matrix elements and the synaptic weights. Next, an implementation of the BAN is presented, showing the possibility to control the output by an external gate voltage.   

\section{Device model and formalism}

\begin{figure}[t]%
	\includegraphics*[width=0.95\linewidth]{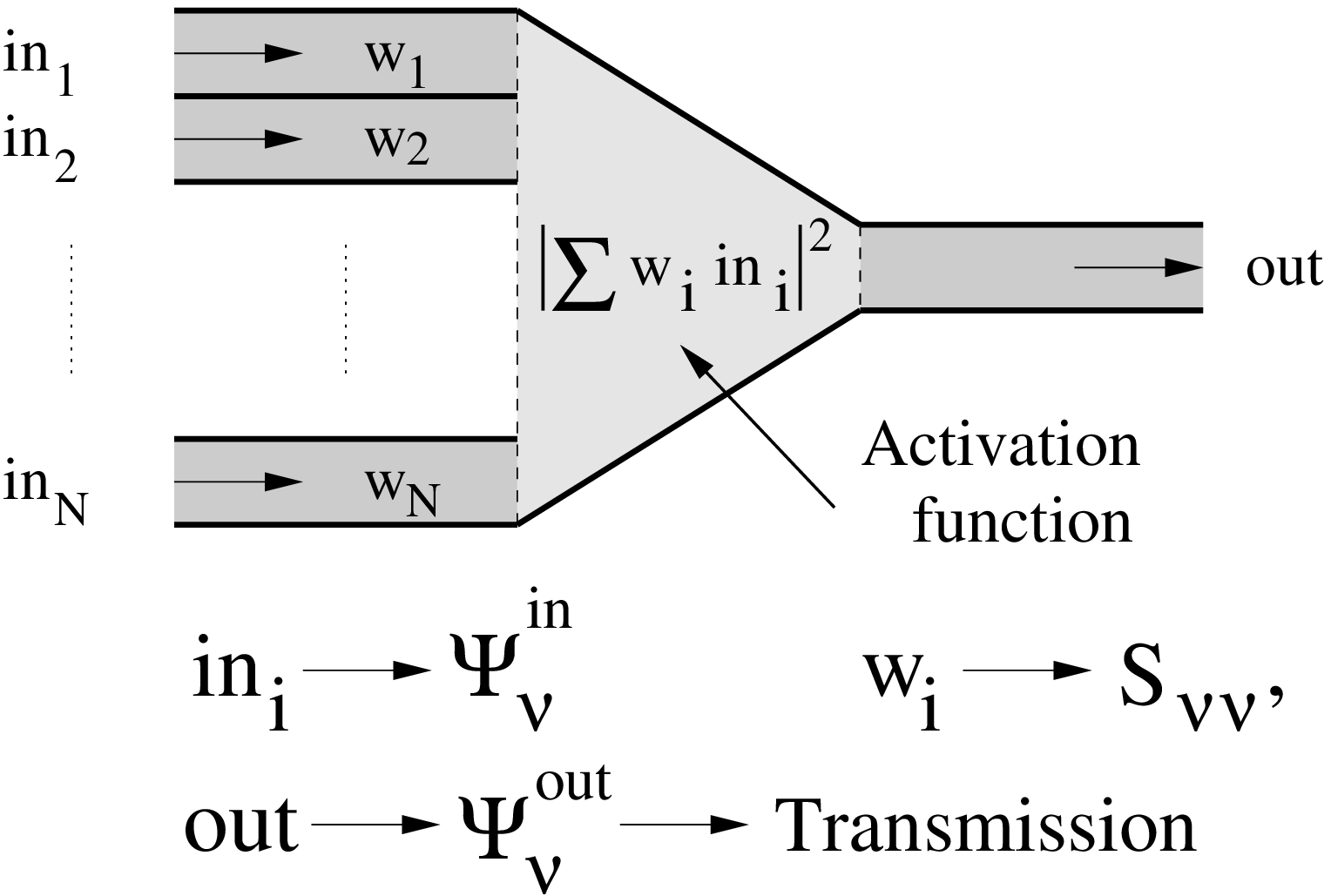}\vspace*{0.2cm}\\
  \includegraphics*[width=0.85\linewidth]{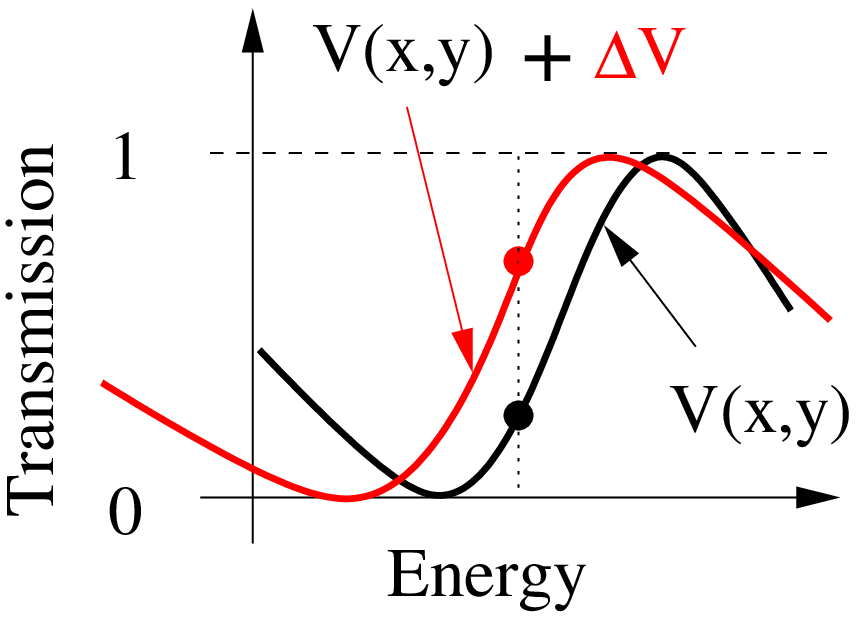}%
	\caption{A schematic representation of the ballistic artificial neuron (upper part): the inputs correspond to incoming  wavefunction coefficients on the left hand side, the synaptic weights ($w_i$) are associated with scattering matrix elements and the output is given by the transmission function given by the outgoing wavefunction coefficients in the right lead. An external gate voltage can tune the transmission function, which is essential for training the ANN (lower part).}
    \label{BAN}
\end{figure}

The BAN model is schematically illustrated in Fig.\ \ref{BAN}. In consists of $N_{\rm in}$ input terminals and one output terminal with a generic scattering region in-between, which mixes the input signals. It is assumed that the input signal is pre-processed by an electron beam splitter so that the incoming wavefunction is located in one ore more terminals on the left hand side, which corresponds to a coherent superposition of propagating modes in different leads. The output signal is collected by measuring the transmission function. 

In the training process of an ANN, the synaptic weights are tuned in order to correct the output so that it gets closer to a target value. In this model, the $\tilde{S}$-matrix elements are adjusted by external gates, which modify the scattering potential.

The coherent transport in the multi-terminal device is described in the framework of a scattering formalism based on the R-matrix method. Given the particularities of the coherent resonant transport, which is typically marked by a rapidly varying transmission function, the R-matrix method efficiently provides the  transmission functions for a relatively large energy set. As usual, the system is divided into leads and scattering region. In the R-matrix method the scattering problem is solved in two steps. In a first step, which is independent on the total energy, the Wigner-Eisenbud problem is solved, corresponding to the solution of the Schr\"odinger equation inside the scattering region with known boundary conditions. This step is most time consuming as it involves the diagonalization of the Hamiltonian. Next, in a second step, the R-matrix, the scattering S-matrix and transmission functions are found for a given energy with significantly smaller effort.   

The scattering problem is formulated as a time independent Schr\"odinger equation with asymptotic boundary conditions, corresponding to particles incoming from one or more leads with total energy E:
\begin{equation}
\label{sch}
{\mathcal H}\Psi(\rr) = E\Psi(\rr),
\end{equation}  
We denote by ${\mathcal H_0}$ the Hamiltonian of the system in the scattering region $\Omega_0$ and by ${\mathcal H}_s$ the Hamiltonians corresponding to the lead regions $\Omega_s$.

The first step includes solving the Wigner-Eisenbud problem, the calculation of the transverse modes and of the overlap integrals. The Wigner-Eisenbud problem is defined for the scattering region:
\begin{equation}
\label{we1}
{\mathcal H_0} \chi_l = E_l \chi_l,\\
\end{equation}
with von Neumann boundary conditions on the frontiers $\Gamma_s = \Omega_s \cap \Omega_0$:	
\begin{equation}	
\label{we2}
\left[ \frac{\partial \chi_l}{\partial z_s} \right]_{\Gamma_s} = 0.
\end{equation}  
The solution yields the Wigner-Eisenbud functions and energies, $\chi_l$ and $E_l$. 

The leads are assumed to be invariant along the transport direction, with a given confinement potential $V_s(\rr^\perp_s)$ so that the transverse modes and energies, $\Phi_\nu$ and $E_{\perp}^{\nu}$, can be readily obtained by solving the Schr\"odinger type equation:
\begin{equation}
{\mathcal H}_s \Phi_\nu = E_{\perp}^{\nu} \Phi_\nu,
\end{equation}  
where $\nu = (s,i)$ is a composite index denoting the terminal index $s$ and the channel index $i$.

Having calculated the Wigner-Eisenbud functions $\chi_l$ and the transverse modes $\Phi_\nu$, one may determine the overlap integrals $(\chi_{l})_\nu$, which are also independent on the total energy $E$:
\begin{equation}
\label{chilnu}
(\chi_{l})_\nu = \int_{\Gamma_s} d \Gamma_s \Phi_\nu(\rr^\perp_s) \chi_l(\rr\in\Gamma_s).
\end{equation}

The wavefunctions inside the leads are superpositions between incoming and outgoing waves so that one may write
\begin{eqnarray}
\label{Psis}
\Psi_s(\rr \in \Omega_s;E)
&=& \sum_i  \Psi_{\nu}^{\rm in} \exp{(-ik_{\nu} z_s)}
     \Phi_\nu(\rr^\perp_s) \nonumber\\
    &+& \sum_i \Psi_{\nu}^{\rm out} \exp{(ik_{\nu} z_s)}
        \Phi_\nu(\rr^\perp_s).
\end{eqnarray}
The wavevectors along the transport direction are $k_{\nu} = \sqrt{ {2m^* \over \hbar^2}(E-E_{\perp}^{\nu}) }$, defined for each channel $\nu$. The channels are open, i.e. propagating waves, if $E_{\perp}^{\nu}<E$ and closed otherwise.

In the second step, the R-matrix is determined for each energy $E$:
\begin{equation}
{\sf R}_{\nu\nu'}(E) = -\frac{\hbar^2}{2} \sum_{l=0}^{\infty}  
                             \frac{(\chi_{l})_\nu(\chi_{l}^*)_{\nu'}}{E-E_{l}}.
\end{equation}
Note that since $(\chi_{l})_\nu$ are known this requires only a summation over the index $l$ corresponding to the Wigner-Eisenbud energies. The cardinality of this set is equal to the basis set used in the Hamiltonian diagonalization.   

The scattering S-matrix, which relates the incoming and outgoing complex amplitudes, $\vec{\Psi}^{\rm out} = S \vec{\Psi}^{\rm in}$, can be determined using the R-matrix:
\begin{equation}
\label{smatrix}
{\sf S} = - \left[ {\sf 1} - \frac{i}{m^*} {\sf R} {\sf k} \right]^{-1} \left[ {\sf 1} + \frac{i}{m^*} {\sf R} {\sf k} \right].
\end{equation}

The current density in lead $s$ averaged in cross-section at $\Gamma_s$ is given by:
\begin{equation}
	\eval{j_s}_{\Gamma_s} = \frac{\hbar}{m^*}\sum_i \left( |\Psi_{\nu}^{\rm in}|^2 - |\Psi_{\nu}^{\rm out}|^2 \right) k_{\nu} .
\end{equation}

Denoting by $\tilde{\Psi}_{\nu}^{\rm in} = \Psi_{\nu}^{\rm in} \sqrt{k_\nu}$ and
$\tilde{\Psi}_{\nu}^{\rm out} = \Psi_{\nu}^{\rm out} \sqrt{k_\nu}$, one may write the current contributions for the incoming and outgoing wavefunctions as:
\begin{eqnarray}
	\eval{j_s^{\rm in}}_{\Gamma_s} &=& \frac{\hbar}{m^*}\sum_i |\tilde{\Psi}_{\nu}^{\rm in}|^2  \\
	\eval{j_s^{\rm out}}_{\Gamma_s} &=& \frac{\hbar}{m^*}\sum_i \left| \sum_{\nu'} \sqrt{\frac{k_\nu}{k_{\nu'}}} S_{\nu\nu'} \tilde{\Psi}_{\nu'}^{\rm in} \right|^2 \\
	&=& \frac{\hbar}{m^*}\sum_i \left| \sum_{\nu'} \tilde{S}_{\nu\nu'} \tilde{\Psi}_{\nu'}^{\rm in} \right|^2,
\end{eqnarray}
where $\tilde{S}$ is the unitary symmetric matrix $\tilde{\sf S}={\sf k}^{1/2}{\sf S}{\sf k}^{-1/2}$ and ${\sf k}$ is a diagonal matrix with matrix elements $k_{\nu'}\delta_{\nu\nu'}$.

The transmission function ${\mathcal T} = j_{\rm R}^{\rm out} / j_{\rm L}^{\rm in}$ is calculated as the ratio between the outgoing particle current in the right terminal and the incoming current in the left terminal.  

Now the correspondence between the elements of McCulloch-Pitts type neurons becomes clear. The inputs correspond to the complex coefficients $\tilde{\Psi}_{\nu}^{\rm in}$ and the synaptic weights are associated with the matrix elements $\tilde{\sf S}_{\nu\nu'}$. If only one outgoing mode is considered, then the standard neuron model is reproduced. Next, by taking the absolute value square a non-linear behavior is introduced as it is usually found by applying typical activation functions in standard ANs.
Note that our proposed ballistic neuron device can be embedded into complex-valued neural networks, which have been investigated in the past years \cite{Hirose:2004:CNN:995119,FAIJULAMIN2009945,dramsch2019complexvalued,scardapane2018complexvalued,Zimmermann2011ComparisonOT,1716132}.

\section{Artificial neuron implementation}

\begin{figure}[t]%
\begin{center}
  \includegraphics[width=0.9\linewidth]{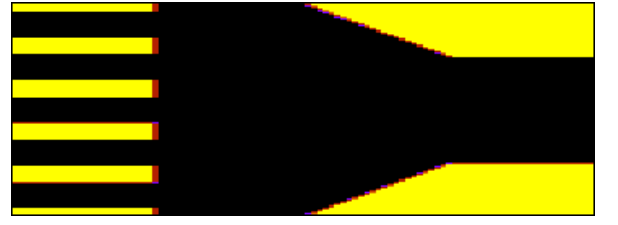}
\end{center}
	\caption{Device scattering potential: black and yellow regions correspond to $V_0$ and $V_b$, respectively.}
    \label{v2D}
\end{figure}

\begin{figure}[t]%
  \includegraphics[width=0.49\linewidth]{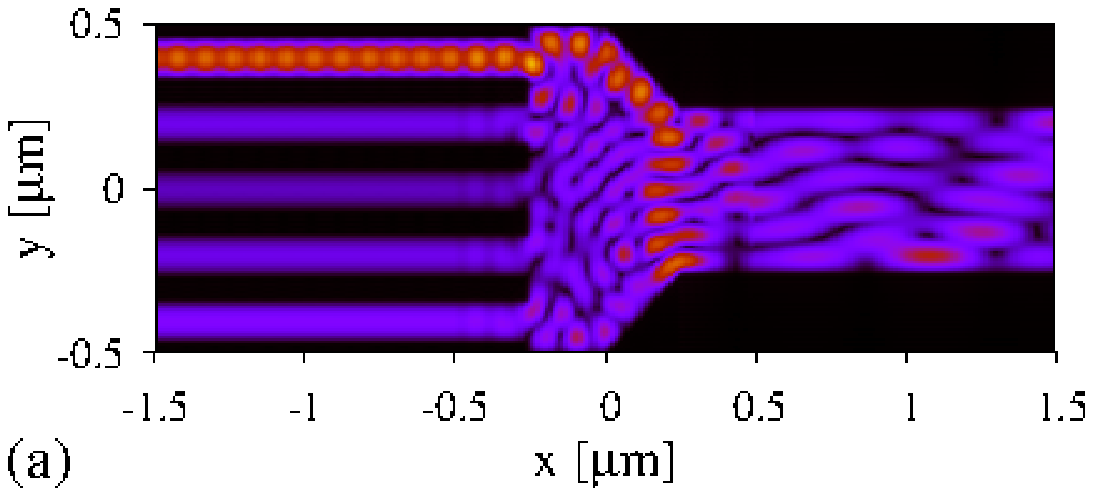}
  \includegraphics[width=0.49\linewidth]{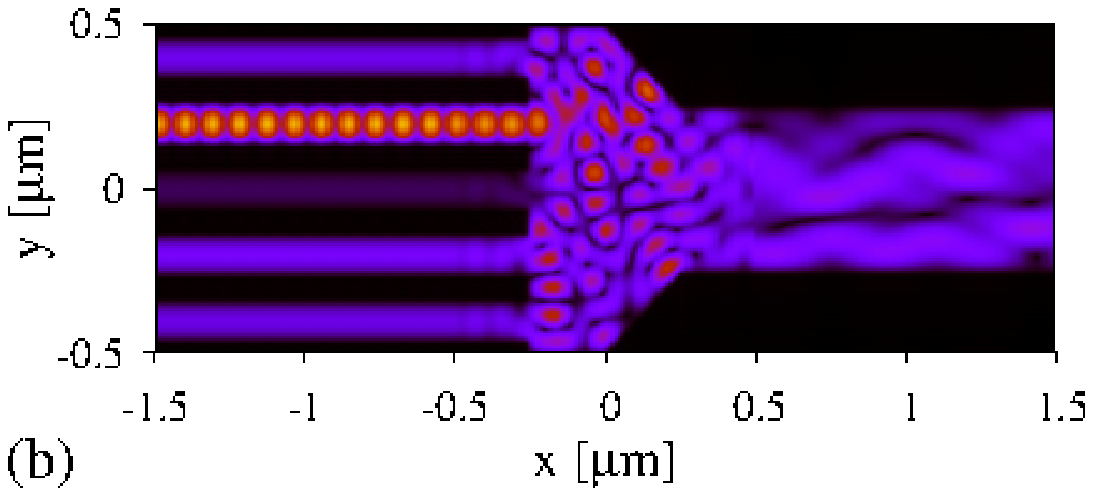}\\
  \includegraphics[width=0.49\linewidth]{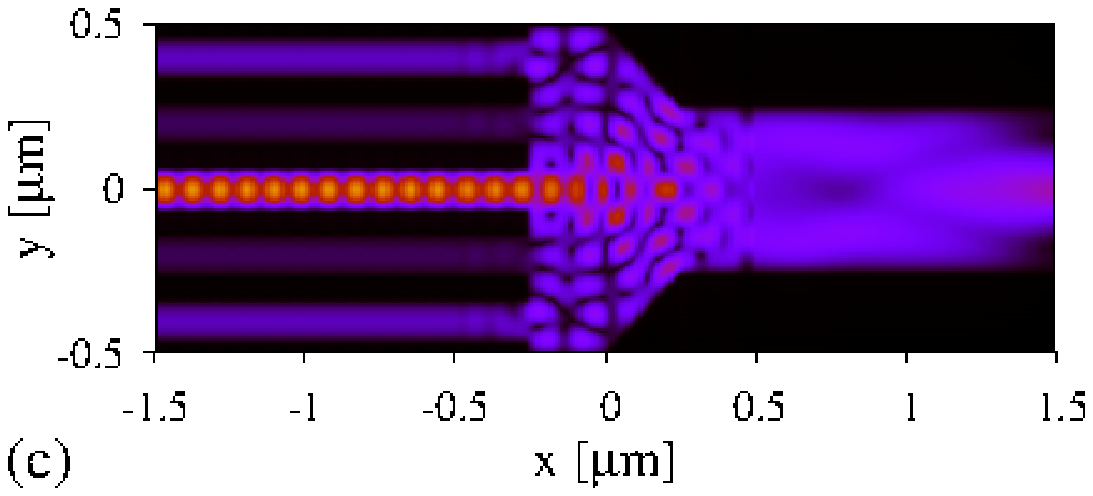}
  \includegraphics[width=0.49\linewidth]{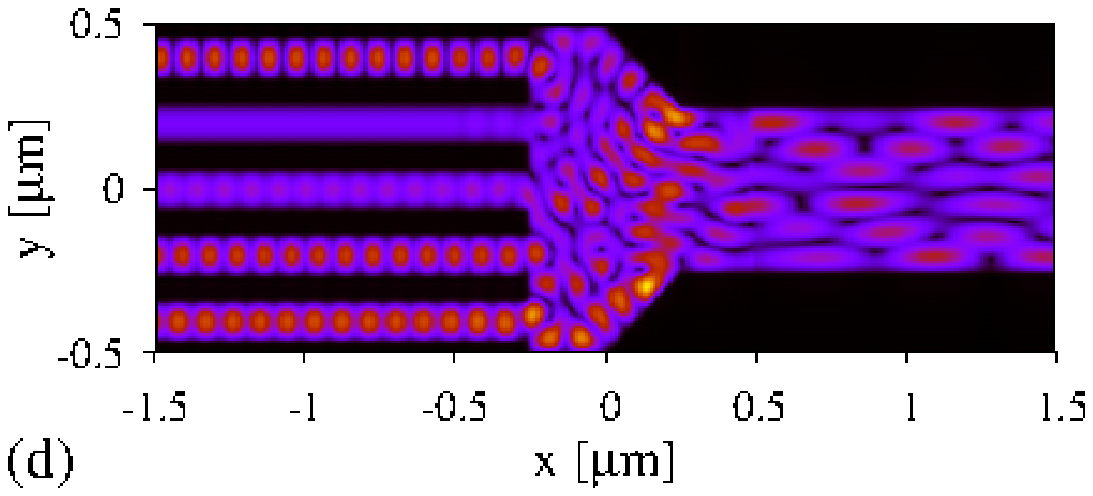}
	\caption{Scattering wavefunctions (absolute value square) corresponding to several input instances set by $\vec{\Psi}^{\rm in}$: (a) $(1,0,0,0,0)$, (b) $(0,1,0,0,0)$, (c) $(0,0,1,0,0)$ and (d) $(1,0,1,\sqrt{2},\sqrt{3})/\sqrt{7}$.}
    \label{wf1}
\end{figure}

The active part of the BAN corresponds to the scattering region $\Omega_0$, which is a rectangular area defined by $(-d_1,d_1)\times(-d_2,d_2)$. On the left hand side there are $N_{\rm in}=5$ input terminals, all having the same width of $w_{\rm in}=d_2/N_{\rm in}$ and one output terminal with a channel width $w_{\rm out}$, which can vary. In order to ensure a smooth connectivity between the leads and the scattering region, the potential inside the leads ($\Omega_s$) is further extended into $\Omega_0$ on a distance of $d_1/2$ on both sides. The potential map contains two values, $V_0$ as reference potential and $V_b$ as the barrier potential which induces electron confinement, as depicted in Fig.\ \ref{v2D}.

For the setup analyzed here the following parameters were considered: $d_1=d_2=500$ nm, $V_0=0$, $V_b=0.1$ meV, and an effective mass $m^*=0.023$ corresponding to InSb. The first transversal eigenmode energy is $\sim$ 1.01 meV and, in the following, we shall consider energies for the incoming electrons small enough to allow only one propagating mode ($E<4$ meV).

Figure\ \ref{wf1} depicts the scattering wavefunctions for an energy of the incident electrons $E=3$ meV. We analyzed the situation where electrons are incoming on single leads $s=0,1,2$ in Fig.\ \ref{wf1}(d) and as a coherent superposition from several leads, e.g. $\vec{\Psi}^{\rm in} = (1,0,1,\sqrt{2},\sqrt{3})/\sqrt{7}$ as a typical input instance. The scattered wavefunctions are reflected back into the input terminals, leading to small ripples, while the transmitted waves are propagating in the outlet terminal, which has a larger width, $w_{\rm out}=d_2/2$, and has 7 propagating channels for the considered energy. The transmission is favored for the $(0,0,1,0,0)$ symmetric input, as there is a larger coupling to the outlet terminal, although this generally depends on the total energy $E$.

\begin{figure}[t]%
  \includegraphics*[width=\linewidth]{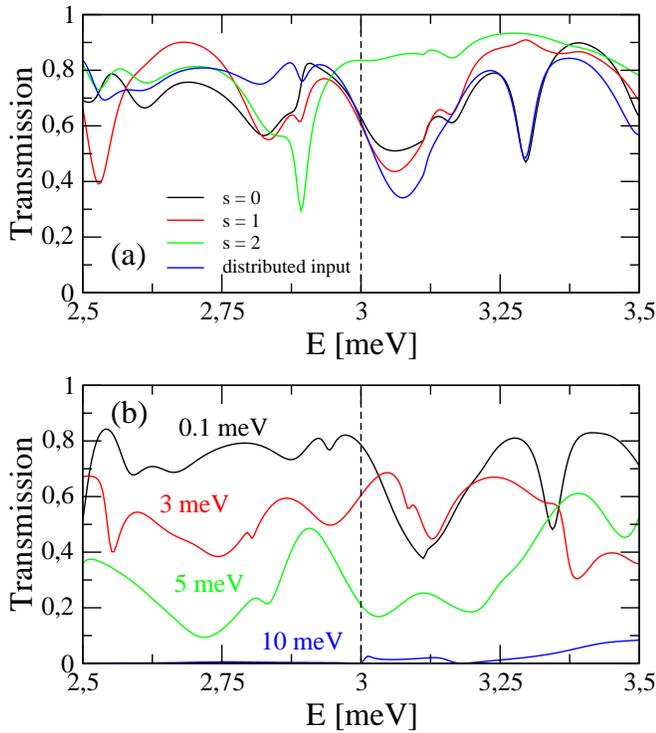}%
	\caption{(a) Transmission functions for the individual and distributed lead inputs considered in Fig.\ \ref{wf1}, for zero gate potential ($V_y = 0$). (b) The tunability of the transmission function is evidenced for the distributed input $(1,0,1,\sqrt{2},\sqrt{3})/\sqrt{7}$ by applying a lateral electric field, as set by $V_y = 0.1, 3, 5, 10$ meV.}
    \label{trans1}
\end{figure}

\begin{figure}[t]%
  \includegraphics[width=0.49\linewidth]{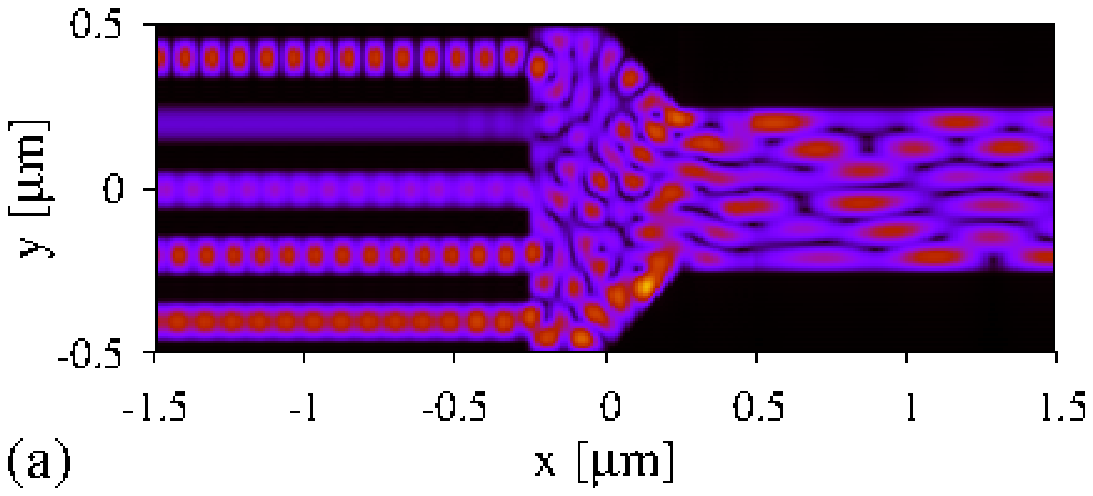}
  \includegraphics[width=0.49\linewidth]{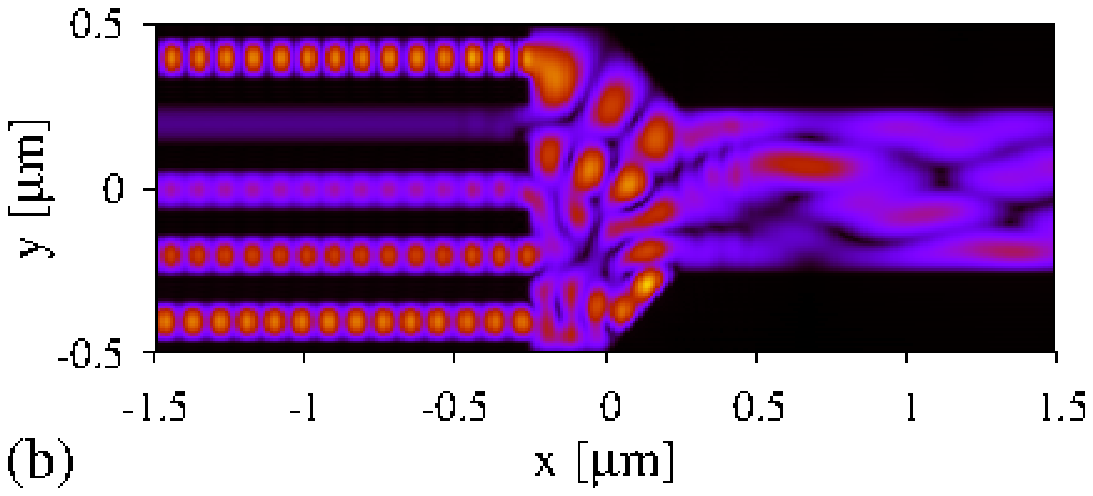}\\
  \includegraphics[width=0.49\linewidth]{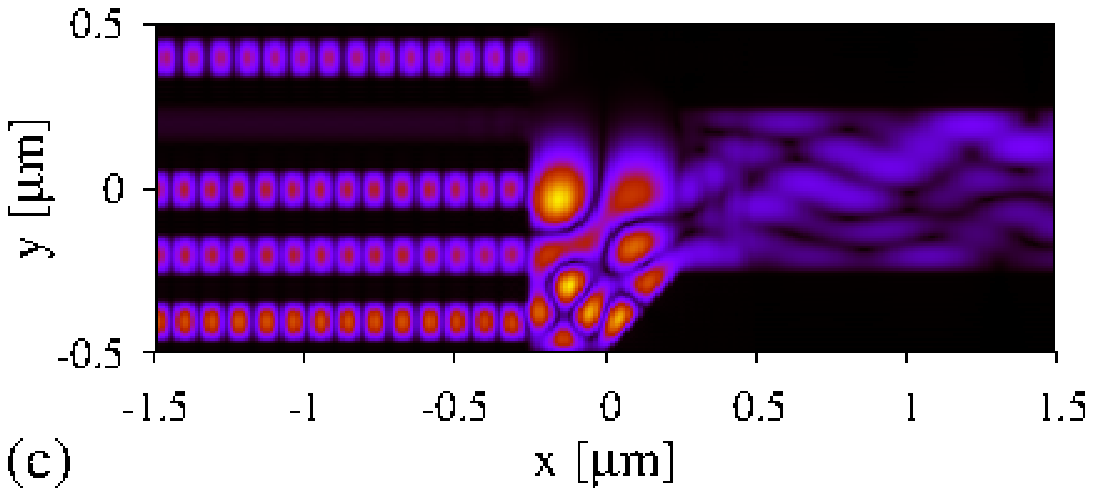}
  \includegraphics[width=0.49\linewidth]{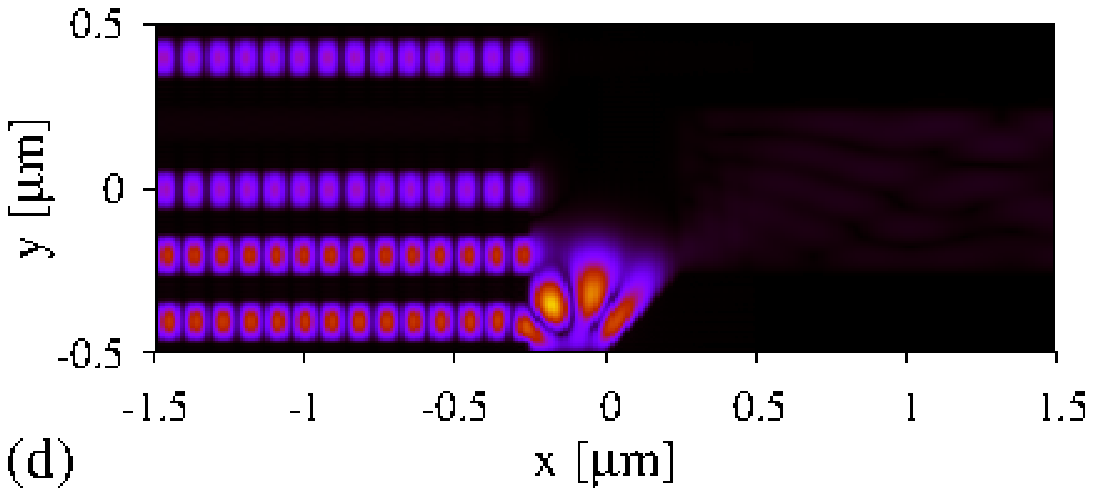}
	\caption{Scattering wavefunctions for an energy $E=3$ meV and different lateral gate fields, specified by $V_y$: (a) 0.1 meV, (b) 3 meV, (c) 5 meV and (d) 10 meV.}
    \label{wf_gf}
\end{figure}

The transmission functions ${\mathcal T}(E)$ for these four input instances are indicated in Fig.\ \ref{trans1}(a) around the value $E=3$ meV. ${\mathcal T}(E)$ presents oscillations, which are characteristic of the resonant transport, with maximum values which are close to 0.9. From the BAN perspective it is important to be able to tune ${\mathcal T}(E)$ int the $(0,1)$ interval by changing the scattering potential, thereby modifying the associated synaptic weights. This can be achieved in several ways. We adopt here a setup with a lateral gate, which changes the potential in the scattering region by $\Delta V(y) = (d_2+y)/(2 d_2) V_y$, for $y\in(-d_2,d_2)$. As $V_y$ is getting larger, some of the input terminals are blocked, reducing the transmission function. This can be seen from Fig.\ \ref{trans1}(b), where $V_y = 0.1, 3, 5, 10$ meV. Blocking barriers can therefore provide a route to decrease the initially high transmission, as shown in Fig.\ \ref{wf_gf}, by analyzing the scattering wavefunctions for $E=3$ meV.

\begin{figure}[h]%
  \includegraphics[width=0.49\linewidth]{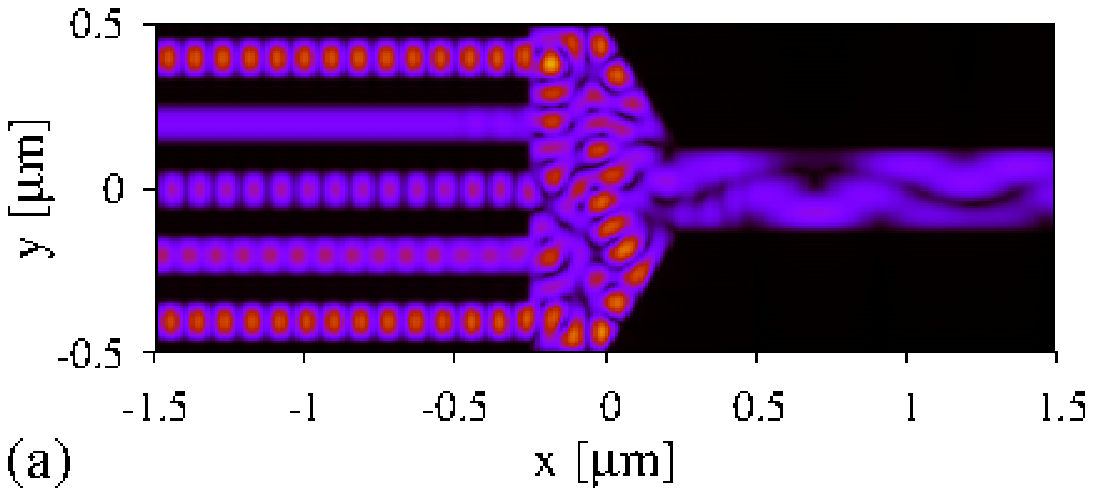}
  \includegraphics[width=0.49\linewidth]{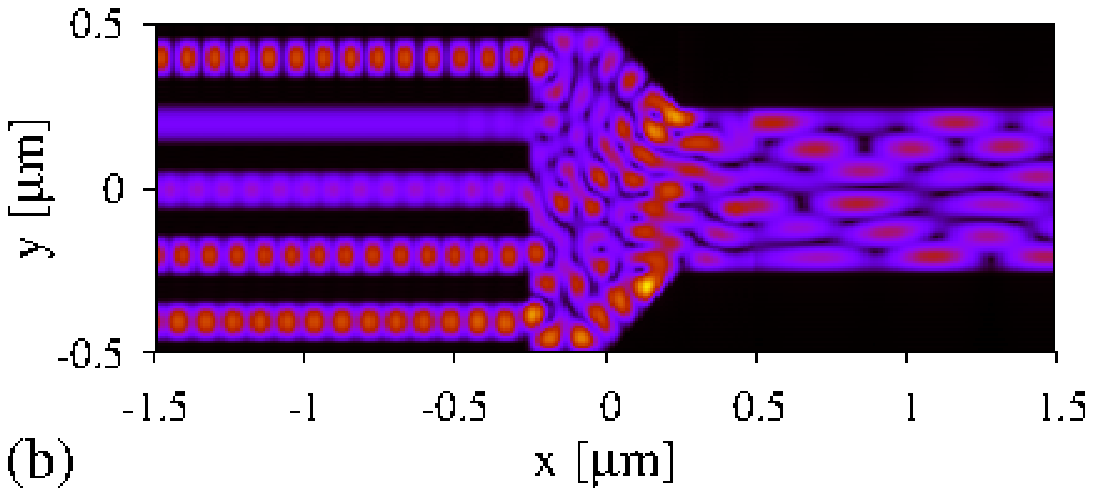}\\
  \includegraphics[width=0.49\linewidth]{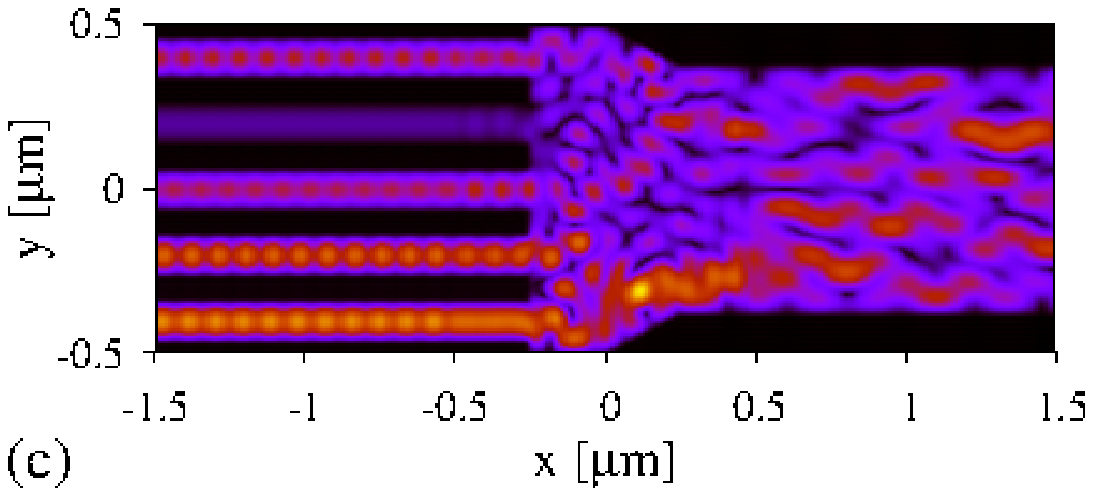}
  \includegraphics[width=0.49\linewidth]{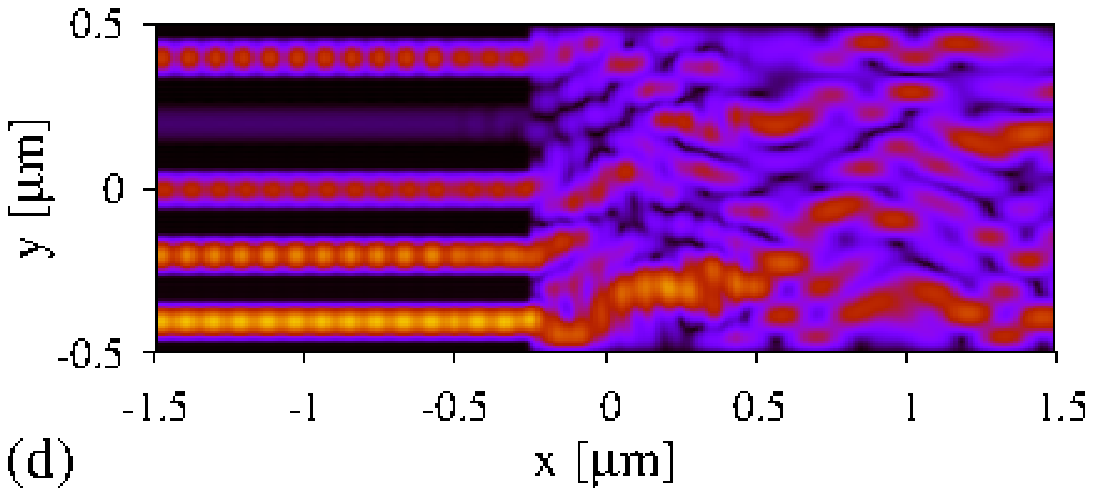}\\
  \includegraphics[width=0.9\linewidth]{figure6e.eps}
	\caption{Scattering wavefunctions for an energy $E=3$ meV and $V_y=0$, by considering different widths of the outlet terminal $w_y/d_2$: (a) 0.25, (b) 0.50, (c) 0.75 and (d) 1. The transmission functions are depicted (e), showing near perfect transmission for the largest width.}
    \label{wf_opw}
\end{figure}

Furthermore, it is also important that high transmission function, close to one, can be achieved for some scattering potential and leads configuration. To this end, we investigated the effect of the outlet terminal width $w_y$ on ${\mathcal T}(E)$ on a broad energy range. From Fig.\ \ref{wf_opw} one can see that ${\mathcal T}(E)$ is significantly enhanced at larger $w_y$, by increasing the number of open channels in the outlet terminal from 4 ($w_y/d_2=0.25$) to 15 ($w_y/d_2=1$). Therefore, the in the training process of an ANN based on the proposed ballistic neurons, one can adjust the synaptic weights by tuning an external gate voltage.


\section{Conclusions}

We introduced a ballistic artificial neuron model based on general scattering matrix theory. The main idea is to include the information regarding the synaptic weights into the device active region instead of storing them in the input terminals. This is achieved by identifying the weights as the scattering matrix elements and the inputs as the incoming wavefunction coefficients. Moreover, taking the absolute value square of the outgoing wavefunctions one obtains the activation function. This configuration allows a compact design, which is essential in terms of power consumption and speed necessary for efficient ANNs. We showed that the output value can be tuned by an external gate potential, which is an important prerequisite for training ANNs. Furthermore, we provided conditions for optimizing the transmission function. The model is applicable in other systems where the S-matrix formalism can be employed, such as optical applications, plasmon polariton systems and microwave network analysis.\\



{\bf Acknowledgements}\\

The authors acknowledge the financial support of project GRAPHENEFERRO grant of Ministry of Research and Innovation, CNCS-UEFISCDI, number PN-III-P4-ID-PCCF-2016-0033.



\bibliography{manuscript}

\end{document}